\documentclass[runningheads]{llncs}

\usepackage[utf8]{inputenc}
\usepackage[T1]{fontenc}
\usepackage[british]{babel}
\usepackage{pbox}
\usepackage{graphicx}
\usepackage{times}
\usepackage{xcolor}
\usepackage{amsmath, amssymb}
\usepackage{dirtytalk}
\usepackage{wrapfig}
\usepackage{listings}
\usepackage[ruled, linesnumbered, commentsnumbered, noend]{algorithm2e}
\usepackage{xspace}
\usepackage[hidelinks]{hyperref}
\usepackage{etoolbox}

\definecolor{bluekeywords}{rgb}{.13, .13, 1}
\definecolor{greencomments}{rgb}{0, .5, 0}
\definecolor{redstrings}{rgb}{.9, 0, 0}
\lstdefinestyle{c}{language=c}
\lstdefinestyle{sum}{numbers=none, basicstyle=\ttfamily\scriptsize}

\SetKwProg{Def}{def}{:}{end}
\SetInd{.3em}{.3em}
\makeatletter
\let\old@algocf@pre@ruled\@algocf@pre@ruled
\renewcommand{\@algocf@pre@ruled}{%
    \Hy@raisedlink{\hyper@anchorstart{algocf.\thealgocf}\hyper@anchorend}%
    \old@algocf@pre@ruled}
\patchcmd{\@algocf@start}{-1.5em}{0pt}{}{}
\makeatother

\newcommand{\LLDD}{\textsc{L2D2}\xspace} 
\newcommand{\Infer}{\textsc{Infer}\xspace}
\newcommand{\CProver}{\textsc{CProver}\xspace}
\newcommand{\RacerX}{\textsc{RacerX}\xspace}

\newcommand{\GoodLock}{\textsc{GoodLock}\xspace}
\newcommand{\AirLock}{\textsc{AirLock}\xspace}
\newcommand{\grep}{\texttt{grep}\xspace}
\newcommand{\sort}{\texttt{sort}\xspace}
\newcommand{\tgrep}{\texttt{tgrep}\xspace}
\newcommand{\find}{\texttt{find}\xspace}
\newcommand{\memcached}{\texttt{memcached}\xspace}
\newcommand{\DDS}{\textsc{Fast-DDS}\xspace}
\newcommand{\eprosimaDDS}{\textsc{eProsima/\DDS}\xspace}
\newcommand{\framac}{\textsc{Frama-C}\xspace}
\newcommand{\mOne}{\texttt{mode\,1}\xspace}
\newcommand{\mTwo}{\texttt{mode\,2}\xspace}

\begin{document}

\newcommand{\thetitle}{Static Deadlock Detection in Low-Level C~Code}
\title{\texorpdfstring{%
    \thetitle\thanks{The work was supported by the project 20-07487S of the
    Czech Science Foundation and the Brno Ph.D. Talent Scholarship
    Programme.}\vspace*{-4mm}%
}{\thetitle}}
\titlerunning{\thetitle}

\author{%
    Dominik Harmim \and
    Vladim\'{\i}r Marcin \and
    Lucie Svobodov\'{a} \and
    Tom\'{a}\v{s} Vojnar\vspace*{-2mm}%
}
\authorrunning{%
    D. Harmim \and
    V. Marcin \and
    L. Svobodov\'{a} \and
    T. Vojnar%
}

\institute{Faculty of Information Technology, Brno University of Technology,
Czech Republic\vspace*{-6mm}}


\hypersetup{
    pdftitle=\thetitle,
    pdfauthor={D. Harmim, V. Marcin, L. Svobodov\'{a}, T. Vojnar},
    pdfsubject={Eurocast'22 Paper},
}

\maketitle

\begin{abstract} We present a~novel scalable deadlock analyser \LLDD capable of
handling C~code with low-level unstructured lock manipulation. \LLDD runs along
the call tree of a program, starting from its leaves, and analyses each function
just once, without any knowledge of the call context. \LLDD builds function
summaries recording information about locks that are assumed or known to be
locked or unlocked at the entry, inside, and at the exit of functions, together
with lock dependencies, and reports warnings about possible deadlocks when
cycles in the lock dependencies are detected. We implemented \LLDD as a~plugin
of the Facebook/Meta \Infer framework and report results of experiments on a
large body of C as well as C++ code illustrating the effectiveness and
efficiency of \LLDD.\end{abstract}

\vspace*{-9mm}\section{Introduction}\vspace*{-2mm}

Nowadays, programs often use \emph{multi-threading} to utilise the many
processors of current computers better. However, concurrency does bring not
only speed-ups but also a~much larger space for nasty errors easy to cause
but difficult to find. The reason why finding errors in concurrent programs
is particularly hard is that concurrently running threads may
\emph{interleave} in many different ways, with bugs hiding in just a~few of
them. Such interleavings are hard to discover by testing even if
it is many times repeated.

Coverage of such rare behaviours can be improved using approaches such
as \emph{systematic testing}~\cite{schedspec12} and \emph{noise-based
testing}~\cite{contestframework03,noise15,anaconda}.
Another way is to use \emph{extrapolating dynamic checkers}, such
as~\cite{fasttrack09,velodrome08}, which can report warnings about possible
errors even if those are not seen in the testing runs, based on spotting
some of their symptoms. Unfortunately, even though such checkers have
proven quite useful in practice, they can, of course, still miss errors.
Moreover, monitoring a~run of large software through such checkers may also
be quite expensive.

On the other hand, approaches based on \emph{model checking}, i.e., exhaustive
state-space exploration, can guarantee the discovery of all potentially
present errors\,---\,either in general or at least up to some bound, which
is usually given in the number of context switches. However, so far, the
scalability of these techniques is not sufficient to handle truly large
industrial code, even when combined with methods such as
\emph{sequentialisation}~\cite{lal-reps-08,lazy-seq-16}, which represents
one of the most scalable approaches in the area.

\enlargethispage{4mm}

An alternative to the above approaches, which can scale better than model
checking and can find bugs not found dynamically (though for the price of
potentially missing some errors and/or producing false alarms), is offered by
approaches based on \emph{static analysis}, e.g., in the form of \emph{abstract
interpretation}~\cite{ai77} or \emph{data-flow analysis}~\cite{dfa73}. The
former approach is supported, e.g., in Facebook/Meta \textsc{Infer}\,---\,an
open-source framework for creating highly scalable, compositional,
incremental, and interprocedural static analysers based on abstract
interpretation~\cite{inferNFM15}.

\Infer provides several analysers that check for various types of bugs, such as
buffer overflows, null-dereferencing, or memory leaks.  However, most
importantly, \Infer is a~\emph{framework} for building new analysers quickly and
easily. As for \emph{concurrency-related bugs}, \Infer provides support for
finding some forms of \emph{data races} and \emph{deadlocks}, but it is limited
to \emph{high-level} Java and C++ programs only and fails for C~programs, which
use a~\emph{lower-level lock manipulation}~\cite{racerD18,inferCACM19}.

In this paper, we propose a~\emph{deadlock checker} that fits the common
principles of analyses used in \Infer and is applicable even to \emph{C~code}
with \emph{lower-level lock manipulation}. Our checker is called \LLDD for
\say{low-level deadlock detector}.

As is common in \Infer, \LLDD computes function summaries \emph{upwards} along
the call tree, starting from its leaves, and analyses every function just once,
without knowing anything about its call contexts.
The summaries contain various pieces of information about locks that are assumed
to be locked/unlocked at the entry of a function, that may be locked/unlocked at
the end of the function, that may be both locked and unlocked inside a function,
as well as about lock dependencies (saying that some lock is locked while
another is still held).
If \LLDD detects a loop in the lock dependencies, it warns about possible
deadlocks.
\LLDD uses multiple heuristics to reduce the number of false alarms, such as
detection of locks serving as gate locks.

To show the effectiveness and efficiency of \LLDD, we present experiments in
which we managed to apply it to 930 programs with 10.3 million lines of code
(MLoC) in total, out of which 8 contained known deadlocks.
\LLDD rediscovered all the deadlocks, and, out of the remaining 922 programs, it
claimed 909 deadlock-free and reported false alarms for 13 of the programs only.
The code included benchmarks coming from the \CProver tool derived from the
Debian GNU/Linux distribution, the code of the \grep, \sort, \tgrep, and
\memcached utilities, and the \eprosimaDDS middleware.

\enlargethispage{4mm}

\vspace*{-3mm}\paragraph{Related Work} 

To the best of our knowledge, \LLDD is the only currently existing, publicly
available, \emph{compositional static deadlock analyser} for \emph{low-level
code}. Below, we briefly discuss approaches that we consider to be the closest
to it.

\RacerX~\cite{racerX03} is a~top-down, non-compositional, flow-sensitive and
context-sensitive analysis for C programs based on computing so-called
\emph{lock sets}, i.e., sets of currently held locks, constructing
a \emph{static lock-order graph}, and reporting possible deadlocks in case of
cycles in it. It employs various heuristics to reduce false-positive reports.
Some of the ideas concerning the lock sets are similar to those used in \LLDD,
and some of the heuristics used in \RacerX inspired those used in \LLDD.

The deadlock analyser implemented within the \CProver
framework~\cite{kroening16} targets C code with POSIX threads and uses a
combination of multiple analyses to create a~context-sensitive and sound
analysis. It also builds a lock-order graph and searches for cycles to detect
deadlocks. Its most costly phase is the pointer analysis used. An experimental
comparison with this tool is given in Section~\ref{sec:experiments}.

\textsc{Starvation}~\cite{deadlock-nikos21} is implemented in the \Infer
framework, and hence it is bottom-up, context-insensitive, and compositional.
It detects deadlocks by deriving lock dependencies for each function and
checking whether some other function uses the locks in an inverse order. It is
thus similar to \LLDD, but \textsc{Starvation} is limited to \emph{high-level}
Java and C++ programs with \emph{balanced locks} only. Moreover, it implements
many heuristics explicitly tailored for Android Java applications.

\GoodLock~\cite{goodlock00} is a~well-known \emph{dynamic analysis} for Java
programs implemented in Java PathFinder (JPF)~\cite{jpf00}. As a representative
of dynamic analysers, it inherits their dependence on the concrete execution (or
executions) of the given software seen for detecting possible deadlocks. It
monitors the lock acquisition history by creating a~\emph{dynamic lock-order
graph}, followed by checking the graph for the existence of deadlock candidates
by searching for cycles in it. To increase chances of spotting even rarely
occurring deadlocks, not directly seen in the given execution, it makes
\emph{deadlock predictions} based on an exponential number of permutations of a
single execution. A~drawback of this approach is that it may produce a high rate
of false positives.

\AirLock~\cite{airlock20} is one of the state-of-the-art dynamic deadlock
analysers. It adopts and improves the basic approach from \GoodLock by applying
various optimisations to the extracted lock-order graph. Moreover, \AirLock,
operating on-the-fly, runs a~polynomial-time algorithm on the lock graph to
eliminate parts without cycles, followed by running a higher-cost algorithm to
detect actual lock cycles.

\vspace*{-3mm}\section{Static Deadlock Detection in Low-Level Concurrent
C~Code}\vspace*{-2mm} \label{sec:l2d2}

This section presents the design of the \LLDD analyser. We first introduce the
main ideas of the analysis, and then discuss it in more detail.

\enlargethispage{6mm}

As already mentioned, \LLDD is designed to handle \emph{C~code} with
\emph{low-level, unstructured lock manipulation}. It does not start the analysis
from the entry code location as done in classical inter-procedural analyses
based, e.g., on~\cite{dfagr95}. Instead, it performs the analysis of a program
function-by-function \emph{along the call tree}, \emph{starting from its
leaves}.
%
%
Therefore, each function is analysed just once without any knowledge of its
possible call contexts. For each analysed function, \LLDD derives a
\emph{summary} that consists of a~\emph{pre-condition} and a
\emph{post-condition}. The summaries are then used when analysing functions
higher up in the call hierarchy. The obtained analysis is \emph{compositional}
on the level of functions, and when used in conjunction with some version
control system, it allows one to focus on \emph{modified functions} and their
dependants only with no need to re-analyse the unchanged functions (which is
typically a vast majority of the code).
%

\begin{wrapfigure}{l}{37mm}
\vspace*{-10mm}
\begin{lstlisting}[
    style=c, label={list:example},
    caption={A sample low-level code causing a deadlock}
]
void f(Lock *<@\textcolor{cyan}{L3'}@>) {
  lock(&<@\textcolor{brown}{L4}@>);
  unlock(&<@\textcolor{cyan}{L3'}@>);
  lock(&<@\textcolor{magenta}{L2}@>);
  <@\ldots@>
  unlock(&<@\textcolor{brown}{L4}@>); }
void *t1(<@\ldots@>) {
  lock(&<@\textcolor{teal}{L1}@>);
  lock(&<@\textcolor{cyan}{L3}@>);
  <@\ldots@>
  f(&<@\textcolor{cyan}{L3}@>);
  unlock(&<@\textcolor{teal}{L1}@>); }
void *t2(<@\ldots@>) {
  lock(&<@\textcolor{magenta}{L2}@>);
  <@\ldots@>
  lock(&<@\textcolor{teal}{L1}@>); }
\end{lstlisting}
\vspace*{-10mm}
\end{wrapfigure}

\LLDD does not perform a classical \emph{alias analysis}, i.e., a precise
analysis for saying whether some pairs of accesses to locks may alias (such an
analysis is considered too expensive\,---\,no such sufficiently precise analysis
works compositionally and at scale).
Instead, \LLDD uses \emph{syntactic access paths}~\cite{ap15}, computed by the
\Infer framework, to represent lock objects. Access paths represent heap
locations via expressions used to access them. In particular, an access path
consists of a base variable followed by a sequence of field selectors.
According to~\cite{racerD18}, the access paths' syntactic equality is a
reasonably efficient way to say (in an under-approximate fashion) that heap
accesses touch the same address. The mechanism is indeed successfully used,
e.g., in the production checker \textsc{RacerD}~\cite{racerD18} to detect data
races in real-world programs.

We will use Listing~\ref{list:example} to illustrate some ideas behind \LLDD.
It works in two phases. In the first phase, it computes a summary for each
function by looking for lock and unlock events (\texttt{lock}/\texttt{unlock}
calls in the listing) in the function. When a call of a user-defined function
appears in the analysed function during the analysis (like on line~11 in the
listing), \LLDD uses a summary of the function if available. Otherwise, the
function is analysed on demand, effectively analysing the code bottom-up (when a
recursive call is encountered, it is skipped). The summary is then applied to an
\emph{abstract state} at the call site. In the listing, the summary
of~\texttt{f} will be applied to the abstract state of~\texttt{t1}.

In the second phase, \LLDD looks through all computed summaries of the analysed
program and focuses on so-called \emph{dependencies} that are a part of the
summaries. These dependencies represent possible locking sequences of the
analysed program. The obtained set of dependencies is interpreted as a relation.
\LLDD computes the transitive closure of this relation and reports a deadlock if
some lock depends on itself in the closure. If we run \LLDD on the code in
Listing~\ref{list:example}, it will report a~potential deadlock due to the
cyclic dependency between the locks \texttt{\textcolor{teal}{L1}} and
\texttt{\textcolor{magenta}{L2}} that arises when the thread \texttt{t1} holds
\texttt{\textcolor{teal}{L1}} and waits on \texttt{\textcolor{magenta}{L2}} and
the thread \texttt{t2} holds \texttt{\textcolor{magenta}{L2}} and waits on
\texttt{\textcolor{teal}{L1}}.

\vspace*{-4mm}\subsection{Computing Function Summaries}\vspace*{-2mm}
\label{sec:summaries}

This section outlines the structure and computation of the summaries used by
\LLDD when analysing some function $f$. Intuitively, the pre-condition expresses
what states of locks $f$ expects from its callers, and the post-condition
reflects the effect of $f$ on the locks. More precisely, the post-condition
includes the \texttt{lockset} and \texttt{unlockset} sets, holding information
about which locks \emph{may be locked} and \emph{unlocked}, resp., at the exit
of $f$. The pre-condition consists of the \texttt{locked} and \texttt{unlocked}
sets, stating which locks are \emph{expected to be locked} and \emph{unlocked},
resp., upon a call of $f$. Note that the \texttt{locked}/\texttt{unlocked} sets
are maintained but not used in the basic algorithm introduced later in this
section. They are used to detect possible
\emph{double-locking}/\emph{unlocking}, see Section~\ref{sec:heuristics}. Next,
the summary's post-condition contains the so-called \emph{lock dependencies}
(\texttt{deps}) in the form of pairs of locks
$(\text{\texttt{\textcolor{magenta}{L2}}, \texttt{\textcolor{teal}{L1}}})$ where
locking of \texttt{\textcolor{teal}{L1}} was observed while
\texttt{\textcolor{magenta}{L2}} was locked. This exact situation can be seen in
Listing~\ref{list:example} on line~16.

\begin{wrapfigure}{l}{.34 \textwidth}
\vspace*{-9mm}
\begin{lstlisting}[
    style=sum, label={list:sum},
    caption={Summaries for the functions from Listing~\ref{list:example}}
]
<@\textbf{f:}@>  <@\textbf{\textcolor{bluekeywords}{PRE-CONDITION}}@>
 locked={<@\textcolor{cyan}{L3'}@>}
 unlocked={<@\textcolor{magenta}{L2}@>, <@\textcolor{brown}{L4}@>}
    <@\textbf{\textcolor{bluekeywords}{POST-CONDITION}}@>
 lockset={<@\textcolor{magenta}{L2}@>}
 unlockset={<@\textcolor{cyan}{L3'}@>, <@\textcolor{brown}{L4}@>}
 wereLocked={<@\textcolor{magenta}{L2}@>, <@\textcolor{brown}{L4}@>}
 deps={(<@\textcolor{brown}{L4}@>, <@\textcolor{magenta}{L2}@>)}
 order={(<@\textcolor{cyan}{L3'}@>, <@\textcolor{magenta}{L2}@>)}
<@\textbf{t1:}@> <@\textbf{\textcolor{bluekeywords}{PRE-CONDITION}}@>
 unlocked={<@\textcolor{teal}{L1}@>, <@\textcolor{magenta}{L2}@>, <@\textcolor{cyan}{L3}@>, <@\textcolor{brown}{L4}@>}
    <@\textbf{\textcolor{bluekeywords}{POST-CONDITION}}@>
 lockset={<@\textcolor{magenta}{L2}@>}
 unlockset={<@\textcolor{teal}{L1}@>, <@\textcolor{cyan}{L3}@>, <@\textcolor{brown}{L4}@>}
 wereLocked={<@\textcolor{teal}{L1}@>, <@\textcolor{magenta}{L2}@>, <@\textcolor{cyan}{L3}@>, <@\textcolor{brown}{L4}@>}
 deps={(<@\textcolor{teal}{L1}@>, <@\textcolor{magenta}{L2}@>), (<@\textcolor{teal}{L1}@>, <@\textcolor{cyan}{L3}@>),
  (<@\textcolor{teal}{L1}@>, <@\textcolor{brown}{L4}@>), (<@\textcolor{cyan}{L3}@>, <@\textcolor{brown}{L4}@>)}
<@\textbf{t2:}@>  <@\textbf{\textcolor{bluekeywords}{PRE-CONDITION}}@>
 unlocked={<@\textcolor{teal}{L1}@>, <@\textcolor{magenta}{L2}@>}
     <@\textbf{\textcolor{bluekeywords}{POST-CONDITION}}@>
 lockset={<@\textcolor{teal}{L1}@>, <@\textcolor{magenta}{L2}@>}
 wereLocked={<@\textcolor{teal}{L1}@>, <@\textcolor{magenta}{L2}@>}
 deps={(<@\textcolor{magenta}{L2}@>, <@\textcolor{teal}{L1}@>)}
\end{lstlisting}
\vspace*{-11mm}
\end{wrapfigure}

Two more sets are a part of the summary's post-condition. First, the
\texttt{wereLocked} set contains information on which \emph{locks may be locked
and then again unlocked} within $f$. This is needed to detect lock dependencies
with such locks in functions higher up in the call hierarchy. Such a situation
can be seen in Listing~\ref{list:example}. The lock
\texttt{\textcolor{brown}{L4}} is locked and then unlocked again within the
function \texttt{f}. In this case, the lock will not be in \texttt{lockset}, and
we would have no information that it was locked there. Consequently, we would
not create any lock dependencies w.r.t. this lock. However, this lock will
appear in \texttt{wereLocked}, so we can create dependencies with it (like the
dependency $(\text{\texttt{\textcolor{teal}{L1}},
\texttt{\textcolor{brown}{L4}}})$ in the function~\texttt{t1} when
calling~\texttt{f} on line~11, which could not be created otherwise).

\enlargethispage{6mm}

The last sets that are a part of \LLDD's post-conditions are denoted as the
\texttt{order} sets. They comprise pairs of locks
$(\text{\texttt{\textcolor{cyan}{L3'}}, \texttt{\textcolor{magenta}{L2}}})$
where locking of~\texttt{\textcolor{magenta}{L2}} was seen when
\texttt{\textcolor{cyan}{L3'}} was unlocked before within the same function.
Such a pair is produced, e.g., on line~4 in Listing~\ref{list:example}. These
sets help \LLDD to better determine the order of operations in functions.
Without it, we would create, e.g., the non-existent dependency
$(\text{\texttt{\textcolor{cyan}{L3}}, \texttt{\textcolor{magenta}{L2}}})$ in
the function \texttt{t1} when calling~\texttt{f} on line~11. It should not be
created because \texttt{\textcolor{cyan}{L3}} is unlocked in \texttt{f} on
line~3 before \texttt{\textcolor{magenta}{L2}} is locked on line~4. Note that
the lock \texttt{\textcolor{cyan}{L3}} from the function \texttt{t1} is passed
to \texttt{f} as \texttt{\textcolor{cyan}{L3'}}. We resolve such situations by
replacing the function's formal parameters with the actual ones at the concrete
call site.

Listing~\ref{list:sum} gives the summaries for the functions in
Listing~\ref{list:example}, omitting the empty sets.

\begin{figure}[t]
\centering
\begin{minipage}{.54 \textwidth}
\centering
\begin{algorithm}[H]
    \KwData{lock~$ L $ being locked; abstract state~$ S $}
    \Def{\texttt{\upshape{lock($ L $, $ S $)}}}{%
        \If{$ L \notin S.locked \cup S.unlocked $}{%
            $ S.unlocked \gets S.unlocked \cup \{L\} $\;
        }
        $ S.lockset \gets S.lockset \cup \{L\} $\;
        $ S.unlockset \gets S.unlockset \setminus \{L\} $\;
        $ S.wereLocked \gets S.wereLocked \cup \{L\} $\;
        $ S.deps \gets S.deps \cup (S.lockset \times \{L\}) $\;
        $ S.order \gets S.order \cup (S.unlockset \times \{L\}) $\;
    }
    \caption{Lock acquisition}
    \label{alg:lock}
\end{algorithm}
\end{minipage}
\hfill
\begin{minipage}{.45 \textwidth}
\centering
\begin{algorithm}[H]
    \KwData{lock~$ L $ being unlocked; abstract state~$ S $}
    \Def{\texttt{\upshape{unlock($ L $, $ S $)}}}{%
        \If{$ L \notin S.locked \cup S.unlocked $}{%
            $ S.locked \gets S.locked \cup \{L\} $\;
        }
        $ S.unlockset \gets S.unlockset \cup \{L\} $\;
        $ S.lockset \gets S.lockset \setminus \{L\} $\;
    }
    \caption{Lock release}
    \label{alg:unlock}
\end{algorithm}
\end{minipage}
\vspace*{-6mm}
\end{figure}

The high-level algorithm for the summary's computation is given in
Algorithms~1--3.
Algorithm~\ref{alg:lock} shows how the abstract state is updated whenever
locking occurs during the analysis. First, it updates the pre-condition by
adding the lock to the \texttt{unlocked} set if this locking is the first
operation with that lock in the given function~$ f $ (lines 2--3). Intuitively,
this reflects that the lock should be unlocked before calling $f$; otherwise, we
would encounter double-locking. Next, the lock acquisition takes place, meaning
that the lock is added to \texttt{lockset} and removed from \texttt{unlockset}
(lines 4--5). Moreover, the lock is added to \texttt{wereLocked} (line~6).
Finally, we derive new dependencies and \texttt{order} edges by considering all
pairs $(L^\prime, L)$ where $L^\prime$ is an element of \texttt{lockset} and
\texttt{unlockset}, resp., and $L$ is the acquired lock (lines 7--8).
Algorithm~\ref{alg:unlock} then updates the abstract state when some lock is
released. It is analogical to the algorithm for locking, but it does not update
the \texttt{wereLocked}, \texttt{deps}, and \texttt{order} sets.

\begin{figure}[t]
\centering
\begin{algorithm}[H]
    \KwData{summary~$ \chi $ of a~callee; abstract state~$ S $}
    \Def{\texttt{\upshape{apply\_summary($ \chi $, $ S $)}}}{%
        $ \chi \gets \mathtt{replace\_formals\_with\_actuals}(\chi) $\;
        \lIf{$ \exists L : L \in \chi.unlocked \wedge L \notin S.unlockset $}{%
            $ S.unlocked \gets S.unlocked \cup \{L\} $%
        }
        \lIf{$ \exists L : L \in \chi.locked \wedge L \notin S.lockset $}{%
            $ S.locked \gets S.locked \cup \{L\} $%
        }
        $ S.lockset \gets (S.lockset \cup \chi.lockset) \setminus
            \chi.unlockset $\;
        $ S.unlockset \gets (S.unlockset \setminus \chi.lockset) \cup
            \chi.unlockset $\;
        $ S.wereLocked \gets S.wereLocked \cup \chi.wereLocked $\;
        $ S.deps \gets S.deps \cup ((S.lockset \times \chi.wereLocked)
            \setminus \chi.order) $\;
    }
    \caption{Integrating a~summary of a~callee}
    \label{alg:sum}
\end{algorithm}
\vspace*{-6mm}
\end{figure}

\enlargethispage{4mm}

Algorithm~\ref{alg:sum} integrates a callee's summary with the abstract
state of an analysed function. Initially, the summary is updated by
replacing the formal parameters with the actual ones (line~2). We also
check that all the locks that should be locked/unlocked before calling the
callee are present in \texttt{lockset}/\texttt{unlockset}, resp. If they
are not, they must be locked/unlocked even before the currently analysed
function. Hence, we update the pre-condition (lines 3--4). On lines 5--7,
the \texttt{lockset}, \texttt{unlockset}, and \texttt{wereLocked} sets are
appropriately modified. At last, new dependencies between the currently
held locks and locks acquired in the callee are introduced (line~8).
However, we exclude all the dependencies from the \texttt{order} set to
avoid adding such $ (L^\prime, L) $ dependencies where~$ L^\prime $ was
unlocked before locking~$ L $ in the callee.

As \LLDD is based on abstract interpretation, we must further define the
\emph{join} operator for combining states along \emph{confluent program paths}
(e.g. in \texttt{if} statements), the \emph{entailment} operator allowing the
analysis to detect it has reached a fixpoint and stop, and the \emph{widening}
operator accelerating the analysis of loops. Since we are interested in locking
patterns along any possible path, we define the join operator as the union of
incoming states' values for all the sets in the summaries. The entailment
operator is defined as testing for a subset on all the sets. The widening
operator is made equal to the join operator as we are working with summaries on
finite and not too large domains.

\vspace*{-4mm}\subsection{Reporting Deadlocks}\vspace*{-2mm}

Checking for deadlocks takes place after the summaries for all functions in the
analysed program are computed. \LLDD then merges all of the derived lock
dependencies into one set $R$. This set is interpreted as a relation, and its
transitive closure $R^+$ is computed. If any lock $L$ depends on itself in the
closure, i.e., $(L, L) \in R^+$, a potential for a deadlock has been detected.
For deadlocks using two locks, \LLDD then looks for dependencies that cause the
deadlock. In particular, it looks for a lock $ L^\prime$ s.t. $(L, L^\prime) \in
R^+ \wedge (L^\prime, L) \in R^+ $ and reports the dependencies (a
generalisation to more locks is, of course, possible).

\vspace*{-2mm}\section{Increasing Analysis Accuracy}\vspace*{-2mm}
\label{sec:heuristics}

\LLDD further implements three heuristics intended to decrease the number of
possible false alarms. We now introduce the two most important (with the third
one being a~simple support for recursive locks).

As \emph{double-locking}/\emph{unlocking} errors are quite rare in practice, the
first heuristic uses their detection as an indication that the analysis is
over-approximating too much. Instead of reporting such errors, \LLDD resets
(some of) the working sets.
%
%
Namely, if a~lock acquisition leads to double-locking, it is assumed that \LLDD
followed some non-existent path, and \texttt{lockset} is no longer trustworthy.
Therefore, it is erased, and the only lock left in it is the currently acquired
one as this is the only one about which we can safely say it is locked. For
that, the following statement is added to Algorithm~\ref{alg:lock}: $\mathbf{if}
\ L \in S.lockset \ \mathbf{then} \ S.lockset \gets \{L\};$.
When releasing a~lock, we then check whether it may already be unlocked. If so,
\texttt{lockset} is erased, eliminating any dependencies that the locking error
would cause. For that, we add the following to Algorithm~\ref{alg:unlock}:
$\mathbf{if} \ L \in S.unlockset \ \mathbf{then} \ S.lockset \gets \emptyset;$.
Finally, we check double-locking/unlocking when a function call is encountered.
We ask whether some lock that should be locked/unlocked in the callee is
currently released/held, resp. If such a lock is found, it is assumed that \LLDD
used a non-existent path to reach the function call, and so \texttt{lockset} is
discarded, and the \texttt{lockset} of the callee will be used instead. We
implement this by adding the following to Algorithm~\ref{alg:sum}: $\mathbf{if}
\ (S.lockset \cap \chi.unlocked \neq \emptyset) \vee$ $(S.unlockset \cap
\chi.locked \neq \emptyset) \ \mathbf{then} \ S.lockset \gets \chi.lockset;$.

\enlargethispage{6mm}

The second heuristic used in \LLDD is the detection of so-called \emph{gate
locks}~\cite{goodlock00}, i.e., locks guarding other locks (upon which deadlocks
on the nested locks are not reported). Whenever we detect a possible
deadlock\,---\,represented by two reverse dependencies $d_1 = (L, L^\prime)$ and
$d_2 = (L^\prime, L)$\,---\,we check whether the same gate lock protects them.
If so, we do not report a deadlock. We check this by computing the intersection
of the \texttt{guards}, i.e., all locks locked before the program points where
the dependencies $d_1$ and $ d_2 $ were captured. In particular, we do not
report a deadlock for dependencies $d_1$ and $d_2$ if
$\text{\texttt{guards}}(d_1) \cap \text{\texttt{guards}}(d_2) \neq \emptyset$.

\vspace*{-4mm}\section{Experimental Evaluation}\vspace*{-2mm}
\label{sec:experiments}

\LLDD has been implemented in OCaml as a plugin of \Infer, and it is publicly
available\footnote{\url{https://github.com/svobodovaLucie/infer}}.
We now report on various experiments we have performed with it.
All of the experiments were run on a machine with the AMD Ryzen 5 5500U CPU,
15\,GiB of RAM, 64-bit Ubuntu 20.04.4 LTS, using \Infer version
v1.1.0-0e7270157.

\begin{wraptable}{r}{65mm}
    \begin{center}
        \vspace*{-14mm}
        \caption{Results of \LLDD and \CProver on non-deadlocking programs of
        the \CProver test-suite}
        \label{tab:resultsDebian}
        \vspace*{1mm}
        \begin{tabular}{l|r|r|r}
        & \textbf{programs} & programs & programs \\
        checker & \textbf{claimed} & raising & failed \\
        & \textbf{safe} & alarms & to analyse \\ \hline
        \textsc{\CProver}  & \textbf{292}   & 114   & 588 \\
        \textsc{\LLDD}\textsubscript{\mOne} & \textbf{906} & 11 & 77 \\
        \textsc{\LLDD}\textsubscript{\mTwo}  & \textbf{896} & 21 & 77
        \end{tabular}
    \end{center}
    \vspace*{-12mm}
\end{wraptable}

In our first set of experiments, we have applied \LLDD on a set of 1,002
C~programs with POSIX threads derived from a Debian GNU/Linux distribution,
originally prepared for evaluating the static deadlock analyser based on the
\CProver framework proposed in~\cite{kroening16}.
The benchmark consists of 11.3\,MLoC.
Eight of the programs contain a~known deadlock.
Like \CProver, \LLDD was able to detect all the deadlocks.
The results for the remaining 994 programs are shown in
Table~\ref{tab:resultsDebian} (for \LLDD, \mOne/\mTwo refer to using/not using
the double-locking-based heuristic), with some more details also in
Table~\ref{tab:detailedResults} discussed below.
We can see that, in \mOne, \LLDD produced 11 false alarms only (77 programs
failed to compile since the \Infer's front-end did not support some of the
constructions used).
We find this very encouraging, considering that the \CProver's deadlock detector
produced 114 false alarms.
Moreover, \LLDD consumed 83 minutes only whereas \CProver needed 4 hours to
handle the programs it correctly analysed, producing 453 timeouts (w.r.t. a
30-minute time limit), and ran out of the available 24\,GB of RAM in 135 cases
(according to \cite{kroening16}, the results were obtained on Xeon X5667 at
3\,GHz running Fedora 20 with 64-bit binaries).

\enlargethispage{6mm}

\begin{wraptable}{r}{76mm}
    \begin{center}
    \vspace*{-13mm}
    \caption{Detailed results on \eprosimaDDS, \sort, \grep, \memcached, \tgrep}
    \vspace*{-2.5mm}
    \label{tab:detailedResults}
    \begin{tabular}{l|r|r|r|r|r}
      & kLoC & \shortstack{ alarms \\ \mOne } & \shortstack{alarms \\ \mTwo} &
        \shortstack{dead- \\ locks} & \shortstack{runtime \\ (mm:ss)} \\ \hline
      \DDS         & 110 & 3 & 6 & 0 & 06:53 \\
      \memcached   & 31 & 6 & 7 & 0 & 00:08 \\
      \sort        & 7.2 & 0 & 0 & 0 & 00:02 \\
      \grep        & 8.7 & 0 & 0 & 0 & 00:03 \\
      \tgrep       & 2.4 & 0 & 0 & 0 & 00:01 \\
      \CProver     & 10,164 & 23 & 80 & 8 & 83:23 \\
    \end{tabular}
    \end{center}
    \label{tab:my_label}
    \vspace*{-11mm}
\end{wraptable}

Table~\ref{tab:detailedResults} provides our further experimental results.
Unlike Table~\ref{tab:resultsDebian}, the table gives not only numbers of
programs in which an alarm was raised, but it gives concrete numbers of the
alarms (more alarms can be raised in a~single program).
Moreover, it shows how \LLDD behaved on multiple further real-life programs.
In particular, \eprosimaDDS 2.6.1 is a C++ implementation of the
Data Distribution Service of the Object Management Group.
For its analysis, we replaced the C++ guard lock used, which is so far not
supported by \LLDD, by a normal lock (exploiting the fact that \Infer
automatically adds all needed \texttt{unlock} calls).
Next, we analysed \memcached version 1.6.10, a distributed memory object caching
system.
The source code of this program was pre-processed by \framac \cite{frama-c},
and we report on the size of the pre-processed code (likewise with all the
further mentioned programs).
Finally, we also analysed \grep 3.7, \tgrep (a multi-threaded version of \find
combined with \grep by Ron Winacott), and GNU Coreutils \sort 8.32.
The alarms raised for \DDS are false alarms caused by some intricacy of C++
locks for which \LLDD was not prepared.
We were not able to check the status of the alarms raised for \memcached, but we
consider them likely false alarms.
However, we find the results provided by \LLDD as quite encouraging since the
numbers of false alarms are low w.r.t. the number of programs and their extent,
and, moreover, we believe that there is space for further improvements
(especially, but not only for C++ locks).

\enlargethispage{6mm}

\paragraph*{Remark} This preprint was submitted for publication in the
post-proceedings of the 18th International Conference on Computer Aided Systems
Theory\,---\,Eurocast'22.
The preprint has not undergone any post-submission improvements or corrections.
The Version of Record of this contribution will be published in LNCS, Springer.




\renewcommand\refname{References\vspace*{-2mm}}
\bibliographystyle{abbrv}
\bibliography{bibliography}

\end{document}